\documentclass[prb,showpacs,twocolumn,amsmath,amssymb,floatfix]{revtex4-1}
\usepackage{graphicx}
\usepackage{bm}
\usepackage{bbm}
\usepackage{color}
\usepackage{footnote}
\usepackage{epstopdf}
\usepackage{hyperref}
\usepackage{upgreek}
\usepackage{appendix}
\usepackage[applemac]{inputenc}

\begin{document}
\title{Andreev spectrum and supercurrents in nanowire-based SNS junctions containing Majorana bound states}
\author{Jorge Cayao,$^{1}$ Annica M. Black-Schaffer,$^{1}$ Elsa Prada,$^2$ Ram\'{o}n Aguado$^3$}
\affiliation{$^1$Department of Physics and Astronomy, Uppsala University, Box 516, S-751 20 Uppsala, Sweden\\$^2$Departamento de F\'{i}sica de la Materia Condensada, Condensed Matter Physics Center (IFIMAC) and Instituto Nicol\'{a}s Cabrera, Universidad Aut\'{o}noma de Madrid, E-28049 Madrid, Spain\\$^3$Instituto de Ciencia de Materiales de Madrid (ICMM-CSIC), Cantoblanco, 28049 Madrid, Spain}
\begin{abstract}
Hybrid superconductor-semiconductor nanowires with Rashba spin-orbit coupling are arguably becoming the leading platform for the search of Majorana bound states (MBSs) in engineered topological superconductors.
We perform a systematic numerical study of the low-energy Andreev spectrum and supercurrents in short and long superconductor-normal-superconductor junctions made of nanowires with strong Rashba spin-orbit coupling, where an external Zeeman field is applied perpendicular to the spin-orbit axis.
In particular, we investigate the detailed  evolution of the Andreev bound states from the trivial into the topological phase and their relation with the emergence of MBSs.  Due to finite length, the system hosts four MBSs, two at the inner part of the junction and two at the outer one. They hybridize and give rise to a finite energy splitting at superconducting phase difference $\phi=\pi$, a well-visible effect that can be traced back to the evolution of the energy spectrum with the Zeeman field: from the trivial phase  with Andreev bound states into the topological phase with MBSs. 
Similarly, we carry out a detailed study of supercurrents for short and long junctions from the trivial to the topological phases.
The supercurrent, calculated from the Andreev spectrum, is $2\pi$-periodic in the trivial and topological phases. In the latter it exhibits a clear \emph{sawtooth} profile at phase difference of $\pi$ when the energy splitting is negligible, signalling a strong dependence of current-phase curves on the length of the superconducting regions.  Effects of temperature, scalar disorder and reduction of normal transmission on supercurrents are also discussed. Further, we identify the individual contribution of MBSs. In short junctions the MBSs determine the current-phase curves, while in long junctions the spectrum above the gap (quasi-continuum) introduces an important contribution.
\end{abstract}
\maketitle

\section{Introduction}
\label{intro}

A semiconducting nanowire with strong Rashba spin-orbit coupling (SOC) 
with proximity-induced $s$-wave superconducting correlations can be tuned into a topological superconductor by means of an external Zeeman field.\cite{PhysRevLett.105.077001,PhysRevLett.105.177002,Alicea:PRB10}
This topological phase is characterized by the emergence of zero-energy quasiparticles with Majorana character localized at the nanowire ends. These Majorana bound states (MBSs) are attracting a great deal of attention owing to their potential for topological, fault-tolerant quantum computation.\cite{kitaev,RevModPhys.80.1083,Sarma:16}
Tunneling into such zero-energy MBSs results in a zero-bias peak of high $2e^2/h$ in the tunnelling conductance in normal-superconductor (NS) junctions  due to perfect Andreev reflection into a particle-hole symmetric state.\cite{Law:PRL09} Early tunnelling experiments in NS junctions \cite{Mourik:S12,xu,Das:NP12,Finck:PRL13,Churchill:PRB13} reported zero-bias peak values much less than the predicted $2e^2/h$. This deviation from the ideal prediction, together with alternative explanations of the zero-bias peak, resulted in controversy regarding the interpretation. Recent experiments have reported significant fabrication improvements and high-quality semiconductor-superconductor interfaces \cite{chang15,Higginbotham,Krogstrup15,zhang16} with an overall improvement on tunnelling data that strongly supports the observation of MBS. \cite{Albrecht16,Deng16,Nichele17,Suominen17,Zhangcondmat17}

Given this experimental state-of-the-art \cite{Aguadoreview17}, new geometries and signatures beyond zero-bias peaks in NS junctions will likely be explored in the near future. Among them, nanowire-based superconductor-normal-superconductor (SNS) junctions are very promising since they are expected to host an exotic fractional $4\pi$-periodic Josephson effect, \cite{kitaev,Fu:PRB09,Kwon:EPJB03} signalling the presence of MBSs in the junction. While this prediction has spurred a great deal of theoretical activity, \cite{Badiane:PRL11,San-Jose:11a,Pikulin:PRB12,PhysRevB.94.085409,PhysRevB.92.134508,Mirceacondmat17,PhysRevB.91.024514,Hansen16} experiments are still scarce, \cite{Woerkom17} arguably due to the lack of good junctions until recently. The situation is now different and, since achieving high-quality interfaces is no longer an issue, Andreev level spectroscopy  and phase-biased supercurrents should provide additional signatures for the unambiguous detection of MBSs in nanowire SNS junctions. Similarly, multiple Andreev reflection transport in voltage-biased SNS junctions \cite{Kjaergaard17,Goffman17} is another promising tool to provide further evidence of MBSs.\cite{SanJoseNJP:13}

Motivated by this, we here present a detailed numerical investigation of the formation of Andreev bound states (ABSs) and their evolution into MBSs in nanowire-based short and long SNS junctions biased by a superconducting phase difference  $\phi$. Armed with this information, we also perform a systematic study of the phase-dependent supercurrents in the short and long junction limits. Due to finite length, the junction always hosts four MBSs in the topological regime: apart from the MBSs located at the junction (inner MBSs), two extra MBSs are located at the nanowire ends (outer MBSs). Despite the early predictions \cite{kitaev,Fu:PRB09,Kwon:EPJB03} of a $4\pi$-periodic Josephson effect in superconducting junctions containing MBSs, in general we demonstrate that the unavoidable overlap of these MBSs renders the equilibrium Josephson effect 2$\pi$-periodic \cite{San-Jose:11a,Pikulin:PRB12} in short and long junctions, since they hybridize either through the normal region or through the superconducting regions giving rise to a finite energy splitting at phase difference $\phi=\pi$. As an example, our calculations show that, for typical InSb parameters, one needs to consider junctions with long superconducting segments of the order of $L_{\rm S}\geq 4\mu$m, where $L_{\rm S}$ is the length of the S regions, in order to have negligible energy splittings.

In particular, we show that in short junctions with $L_{\rm N}\ll \xi$, where $L_{\rm N}$ is the normal region length and $\xi$ is the superconducting coherence length, the four MBSs (\emph{inner} and \emph{outer}) are the only levels within the induced gap. On the contrary, the four MBSs coexist with additional levels in long junctions with  $L_{\rm N}\gg \xi$, which affect their phase dependence.   
Despite this difference, we demonstrate that the supercurrents in both limits exhibits a clear \emph{sawtooth} profile when the energy splitting near $\phi=\pi$ is small, therefore signalling the presence of weakly overlapping MBSs.  We find that while this sawtooth profile is robust against variations in the normal transmission and scalar disorder, it smooths out when temperature effects are included, making it a fragile, yet useful, signature of MBSs. 

We identify that in short junctions the current-phase curves are mainly determined by the levels within the gap, the four MBSs, with only very little quasi-continuum contribution. In long junctions, however, all the levels within the gap, MBSs and the additional levels due to longer normal region together with the quasi-continuum determine the current-phase curves. In this situation, the additional levels that arise within the gap disperse almost linearly with $\phi$ and therefore affect the features of the supercurrents carried by MBSs only.

Another important feature we find is that the current-phase curves do not depend on $L_{\rm S}$ in the trivial phase (for both short and long junctions), while they strongly depend on $L_{\rm S}$ in the topological phase. Our results demonstrate that this effect is purely connected to the splitting of MBSs at $\phi=\pi$, indicating another unique feature connected with the presence of MBSs in the junction. In fact, the maximum of such current-phase curves in the topological phase increases as the splitting is reduced, saturating when the splitting is completely suppressed. This, together with the sawtooth profile in current-phase curves, correspond to the main findings of this work.
Results presented here therefore strongly complement our previous study on critical currents \cite{Cayao17b} and should provide useful insight for future experiments looking for Majorana-based signatures in nanowire-based SNS junctions.

The paper is organised as follows. In Sec.\,\ref{sec0} we describe the model for semiconducting nanowires with SOC, where we show that only the right combination of Rashba SOC, a Zeeman field perpendicular to the spin-orbit axis and $s$-wave superconductivity leads to the emergence of MBSs. Similar results have been presented elsewhere but we include them here for the sake of readability of the next sections. In Sec.\,\ref{sec1} we discuss how nanowire-based SNS junctions can be readily modelled using the tools of Sec.\,\ref{sec0}. Then, we describe the low-energy Andreev level spectrum and its evolution from the trivial into the topological phase with the emergence of MBSs. 
In the same section, we report results on the supercurrent which exhibit a \emph{sawtooth} profile at $\phi=\pi$ as a signature of the emergence of MBSs. In Sec.\,\ref{concl} we present our conclusions. For completeness, in App.\,\ref{AppA} we  show localization and  exponential decay as well as homogeneous oscillations of the MBSs in wires and SNS junctions. Additional calculations of the Zeeman dependent low-energy spectrum in SNS junctions are also presented.

\section{Nanowire model}
\label{sec0}
The aim of this part is to properly describe the emergence of MBSs in semiconducting nanowires with SOC. 
We consider a single channel nanowire in one-dimension with SOC and Zeeman interactions, whose model Hamiltonian is given by\cite{rashba84a,rashba84b,PhysRevB.66.073311,PhysRevLett.90.256601,SOCExp10,PhysRevB.97.035433}
\begin{equation}
\label{H0Hamil}
H_{0}=\frac{p^{2}_{x}}{2m}-\mu-\frac{\alpha_{\rm R}}{\hbar}\sigma_{y}p_{x}+B\sigma_{x}\,,
\end{equation}
where $p_{x}=-i\hbar\partial_{x}$ is the momentum operator, $\mu$ the chemical potential which determines the filling of the nanowire, $\alpha_{\rm R}$ represents the strength of Rashba spin-orbit coupling, $B=g\mu_{\rm B}\mathcal{B}/2$ is the Zeeman energy as a result of the applied magnetic field $\mathcal{B}$ in the x-direction along the wire, $g$ is the wire's $g$-factor and $\mu_{\rm B}$ the Bohr magneton. Parameters for InSb nanowires include \cite{Mourik:S12} include the electron's effective mass $m=0.015m_{e}$, with $m_{e}$ the electron's mass, and the spin-orbit strength $\alpha_{R}=20$\,meVnm. 

\begin{figure}
\centering
\includegraphics[width=.45\textwidth]{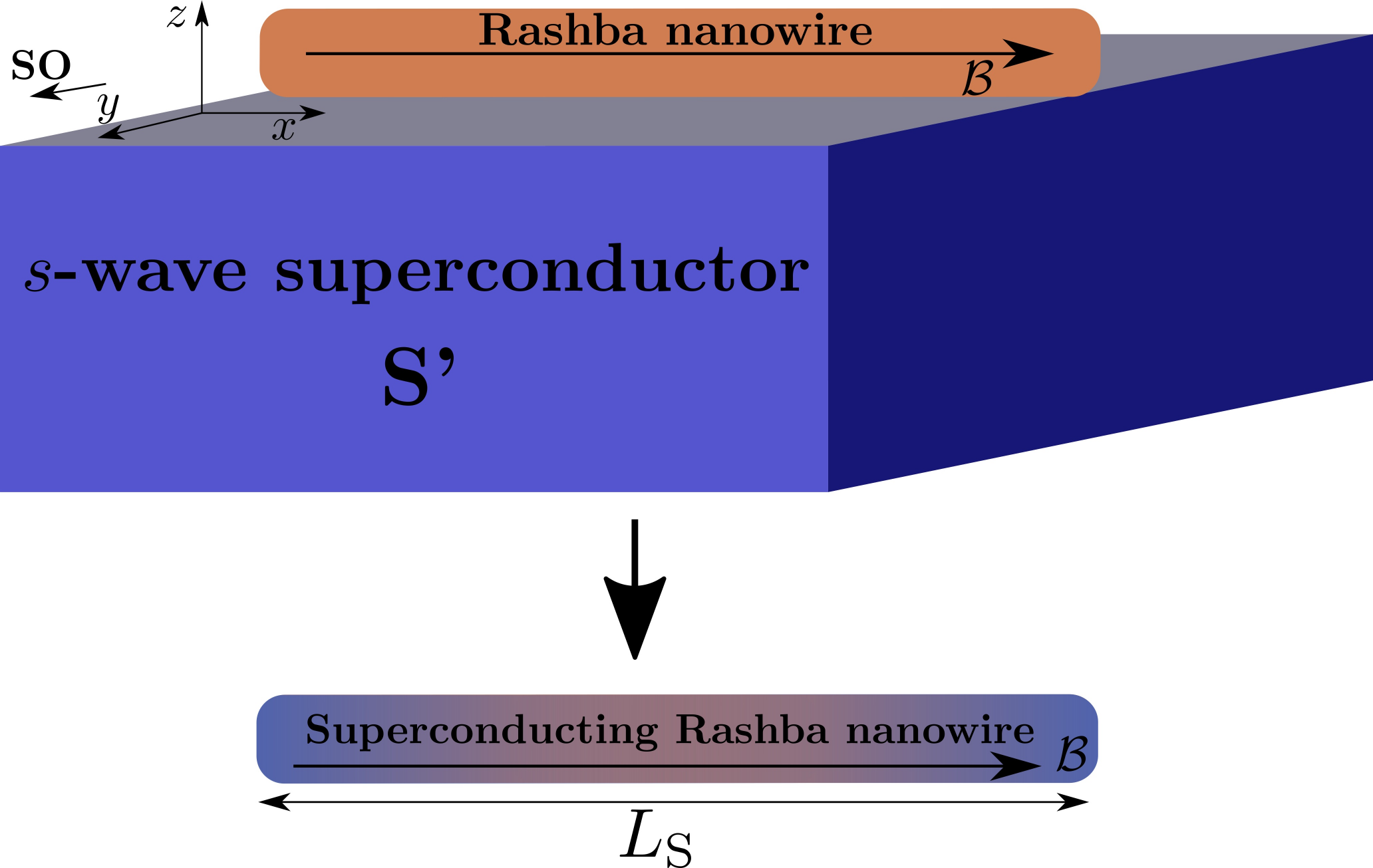} 
\caption[Rashba nanowire]{(Color online) A semiconducting nanowire with Rashba SOC is placed on a $s$-wave superconductor (S') with pairing potential $\Delta_{S'}$ and it is subjected to an external magnetic field $\mathcal{B}$ (denoted by the black arrow). 
Superconducting correlations are induced into the nanowire via proximity effect, thus becoming superconducting with induced pairing potential 
$\Delta_{\rm S}<\Delta_{S'}$.}
\label{Fig0}
\end{figure}

We consider a semiconducting nanowire placed in contact with an $s$-wave superconductor with pairing potential $\Delta_{\rm S'}$ (which is in general complex) as schematically shown in Fig.\,\ref{Fig0}. Electrons in such nanowire feel an effective superconducting pairing potential as a result of the so-called \emph{proximity effect}.\cite{RevModPhys.36.225,Doh272} In order to have a good proximity effect, a highly transmissive interface between the nanowire and the superconductor is required, so that electrons can tunnel between these two systems.\cite{chang15,Higginbotham,Krogstrup15,zhang16} This results in a superconducting nanowire, with a well defined induced hard gap (namely, without residual quasiparticle density of states inside the induced superconducting gap). The model describing such proximitized nanowire can be written in the basis  $(\psi_{\uparrow},\psi_{\downarrow},\psi^{\dagger}_{\uparrow},\psi^{\dagger}_{\downarrow})$ as
\begin{equation}
\label{SCNW}
H=
\begin{pmatrix}
H_{0}&\Delta_{S}(x)\\
\Delta^{\dagger}_{S}(x)&-H_{0}^{*}
\end{pmatrix}\,,
\end{equation}
where $\Delta_{\rm S}<\Delta_{\rm S'}$. Since superconducting correlations are of $s$-wave type, the pairing potential is given by
\begin{equation}
\Delta_{S}(x)=i\sigma_{y}\,\Delta\,{\rm e}^{i\phi}\,,
\end{equation}
where $\phi$ is the superconducting phase. We set $\phi=0$ when discussing superconducting nanowires, while the SNS geometry of course allows a finite phase difference $\phi\neq 0$ across the junction.

It was shown \cite{PhysRevLett.105.177002,PhysRevLett.105.077001,Alicea:RPP12} that the nanowire with Rashba SOC and in proximity to an $s$-wave superconductor, described by  Eq.\,(\ref{SCNW}), contains a topological phase characterized by the emergence of MBSs localized at the wire's ends. This can be understood as follows. The interplay of all these ingredients generates two 
intraband $p$-wave pairing order parameters $\Delta_{--,++}(k)=(\pm\alpha_{R}k\Delta)/\sqrt{B^{2}+\alpha_{R}^{2}k^{2}}$ 
and one interband $s$-wave $\Delta_{+-}=(B\Delta)/\sqrt{B^{2}+\alpha_{R}^{2}k^{2}}$, where $+$ and $-$ denote the Rashba bands of $H_{0}$. 
The gaps associated to the $\pm$  BdG spectrum are different and correspond to the inner and outer part of the spectrum, denoted by  $\Delta_{1,2}$ at low and high momentum, respectively. These gaps  depend in a different way on the Zeeman field. Indeed, as the Zeeman field $B$ increases the gap $\Delta_{1}$, referred to as the inner gap, is reduced
 while $\Delta_{2}$, referred to as the outer gap, is slightly reduced although for strong SOC it remains roughly constant.
The inner gap $\Delta_{1}$ closes at $B=B_{\rm c}$ and reopens for $B>B_{\rm c}$ giving rise to the topological phase, while the outer gap 
remains finite. 
The topological phase is effectively reached due to the generation of an effective $p$-wave superconductor, which is the result of projecting the system 
Hamiltonian onto the lower band ($-$) keeping 
only the intraband $p$-wave pairing 
$\Delta_{--}$.\cite{PhysRevLett.105.077001,PhysRevLett.105.177002} Deep in the topological phase $B>B_{\rm c}$, the lowest gap is $\Delta_{2}$. 

In order to elucidate and visualise the topological transition, we first analyse the low-energy spectrum  of the superconducting nanowire. 
This spectrum can be numerically obtained by discretising the Hamiltonian given by Eq.\,(\ref{H0Hamil}) into a tight-binding lattice 
\begin{equation}
\label{HamilTB}
H_{0}\,=\,\sum_{i}c_{i}^{\dagger}\,h\,c_{i}\,+\,\sum_{<ij>}c_{i}^{\dagger}\,v\,c_{j}\,+\,\text{h.c.}\,,
\end{equation}
where the symbol $<ij>$ means that $v$ couples nearest-neighbor $i,j$ sites; 
$h=(2t-\mu)\sigma_{0}+B\sigma_{x}$ and $v=-t\sigma_{0}+it_{\rm SO}\sigma_{y}$ are matrices in spin space, $t=\hbar^{2}/(2m^{*}a^{2})$ is the hopping parameter and $t_{\rm SOC}=\alpha_{\rm R}/(2a)$ the SOC hopping.  The dimension of $H_{0}$ is set by the number of sites of the wire. Then, it is written in Nambu space as given by Eq.\,(\ref{SCNW}). 
Such Hamiltonian is then diagonalized numerically with its dimensions given by the number of sites $N_{\rm S}$ of the wire. Since this description accounts for finite length wires, it is appropriate for investigating the overlap of MBSs. The length of the superconducting wire is $L_{\rm S}=N_{\rm S}a$, where $N_{\rm S}$ is the number of sites and $a$ the lattice spacing.  As we mentioned, the superconducting phase in the order parameter is assumed to be zero as it is only relevant when investigating Andreev bound states in SNS junctions.

 \begin{figure}
\centering
\includegraphics[width=.45\textwidth]{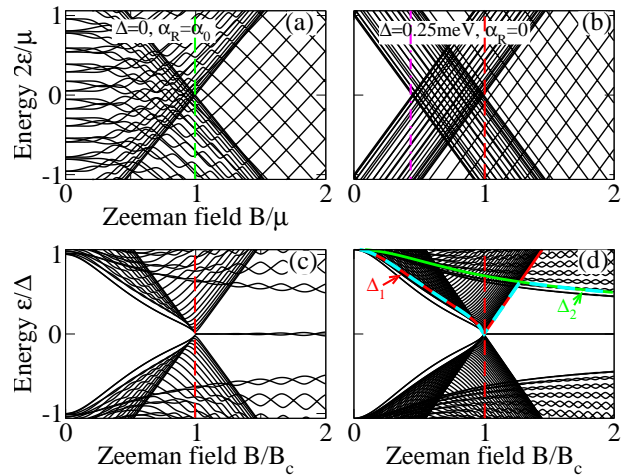} 
\caption[Energy levels as function of the Zeeman field]{(Color online) 
Low-energy spectrum of a superconducting nanowire as function of the Zeeman field $B$.  (a) At zero superconducting pairing with finite SOC  the spectrum is gapless and becomes spin polarized at $B=\mu$ as indicated by the green dashed line, (b) while a finite superconducting pairing with zero SOC induces a gap for low values of $B$.  As $B$ increases, the induced gap is reduced and closed at $B=\Delta$ (vertical magenta dash-dot line).
Bottom panels (c-d) correspond to  both finite superconducting
pairing and SOC for (c) $L_{\rm S}=4000$\,nm and (d) $L_{\rm S}=10000$\,nm.
Notice that as the Zeeman field increases the spectrum exhibits the closing of the gap at $B=B_{\rm c}$. While in the trivial phase, $B<B_{\rm c}$, there are no levels within the reduced gap (bottom panels), in the topological phase, $B>B_{\rm c}$, the two lowest levels develop and oscillatory behaviour around zero energy  (c). These lowest levels are the celebrated MBSs.
For long enough wires (d)  the amplitude of the oscillations is reduced and such levels acquire zero energy. Solid red, green and dashed cyan curves indicate the induced gaps $\Delta_{1,2}$ and ${\rm min}(\Delta_{1},\Delta_{2})$.
Parameters: $\alpha_{0}=20$\,meVnm, $\mu=0.5$\,meV and $\Delta=0.25$\,meV and $L_{\rm S}=4000$\,nm (for top panels).}
\label{SC3}
\end{figure}

In Fig.\,\ref{SC3} we present the low-energy spectrum for a superconducting nanowire as a function of the Zeeman field at fixed wire's length  $L_{\rm S}$.
Fig.\,\ref{SC3}(a) shows the case of zero superconducting pairing and finite SOC ($\Delta=0$, $\alpha_{R}\neq0$), while Fig.\,\ref{SC3}(b) a situation of 
finite pairing but with zero SOC ($\Delta\neq0$, $\alpha_{\rm R}=0$).  These two extreme cases are very helpful in order to understand how a topological transition occurs when the missing ingredient (either superconducting pairing of finite SO) is included. This is illustrated in the bottom panels, which correspond to both finite SOC and superconducting pairing for $L_{\rm S}<2\xi_{\rm M}$ and $L_{\rm S}>2\xi_{\rm M}$, respectively. Here, $\xi_{\rm M}$ represents the Majorana localisation length, which is obtained after solving numerically the polynomial equation \cite{PhysRevLett.105.077001,PhysRevB.91.024514}  $k^{2}+4(\mu+C\alpha_{R}^{2})Ck^{2}+8\lambda C^{2}\Delta \alpha_{R} k+4C_{0}C^{2}=0$, obtained from Eq.\,(\ref{SCNW}), where $C=m/\hbar^{2}$ and $C_{0}=\mu^{2}+\Delta^{2}-B^{2}$. We define the Majorana localization length as $\xi_{\rm M}={\rm max}[-1/k_{sol}]$, where $\{k_{sol}\}$ are solutions to the previous polynomial equation.

For the sake of the explanation, we plot the spectrum in the normal state ($\Delta=0$),  Fig.\,\ref{SC3}(a), which is, of course, gapless. As the Zeeman field increases, the energy levels split and, within the weak Zeeman phase, $B<\mu$, the spectrum contains energy levels with both spin components.  In the strong Zeeman phase, $B>\mu$, one spin sector is completely removed giving rise to a spin-polarised spectrum at low energies as one can indeed observe in Fig.\,\ref{SC3}(a). The transition point from weak to strong Zeeman phases is marked by the chemical potential $B=\mu$ (green dashed line).
Fig.\,\ref{SC3}(b) shows the low-energy spectrum at finite superconducting pairing, $\Delta\neq0$, and zero SOC, $\alpha_{R}=0$.
Firstly, we notice, in comparison with Fig.\,\ref{SC3}(a), that the superconducting pairing induces a gap with no levels for energies below $\Delta$ at $B=0$. This is in accordance with the Anderson's 
theorem which prevents the existence of bound states inside the gap of an $s$-wave superconductor for non-magnetic impurities.\cite{Anderson-theorem} A finite magnetic field induces a so-called \emph{Zeeman depairing}, 
which results in a complete closing of the induced superconducting gap when $B$ exceeds $\Delta$.  
 This is indeed observed in Fig.\,\ref{SC3}(b) (magenta dash-dot line). 
Further increasing of the Zeeman field in this normal state gives rise to a region for $\Delta<B<B_{\rm c}$ which depends on the finite value of the chemical potential (between red and magenta lines) where the energy levels contain both spin components (for $\mu=0$ magenta dash-dot  and  red dashed lines coincide, not shown). Notice that $B_{\rm c}=\sqrt{\Delta^{2}+\mu^{2}}$. On the other hand, for $B>B_{\rm c}$, one spin sector is removed and the energy levels are spin polarised, giving rise to a set of Zeeman crossings which are not protected. Remarkably, when $\alpha_{R}\neq0$, the low-energy spectrum undergoes a number of important changes , Fig.\,\ref{SC3}(c-d).
First, the gap closing changes from $\Delta$, Fig.\,\ref{SC3}(b), to $B_{\rm c}=\sqrt{\Delta^{2}+\mu^{2}}$ (bottom panels). 
Second, a clear closing of the induced gap at $B=B_{\rm c}$ and reopening for $B>B_{\rm c}$ is observed as the Zeeman field increases. This can be seen by plotting the induced gaps $\Delta_{1,2}$, which are finite only at finite Zeeman field. 
In Fig.\,\ref{SC3}(d) red, green and dashed cyan curves correspond to $\Delta_{1}$, $\Delta_{2}$ and ${\rm min}(\Delta_{1},\Delta_{2})$. Remarkably, the closing and reopening of the induced gap in the spectrum exactly (up to some finite size corrections) follows the gaps $\Delta_{1,2}$ derived from the continuum.   
Third, the spin polarised energy spectrum shown in Fig.\,\ref{SC3}(b) at zero SOC for $B>B_{\rm c}$ is washed out, keeping only the crossings around zero energy of the two lowest levels. Such closing and reopening of the spectrum at the critical field $B_{\rm c}$ signals a topological transition where the two remaining lowest energy levels for $B>B_{\rm c}$ are the well-known MBSs. 
Owing to the finite length $L_{\rm S}$, the MBSs exhibit the expected oscillatory behaviour due to their finite spatial overlap.\cite{PhysRevB.86.180503,DasSarma:PRB12,PhysRevB.86.085408,Rainis:PRB13} For long enough wires $L_{\rm S}\gg2\xi_{\rm M}$, the amplitude of the oscillations is considerable reduced (even negligible), which pins the MBSs to zero energy. 
Fourth, the SOC introduces a finite energy separation  between the two lowest levels (crossings around zero) and the rest of the low-energy spectrum denoted here as \emph{topological minigap}.
Notice that the value of this minigap, related to the high momentum gap $\Delta_{2}$, remains finite and roughly constant for strong SOC. In the case of weak SOC the minigap is reduced and for high Zeeman field it might acquire very small values, affecting the topological protection of the MBSs.

To complement this introductory part, calculations of the wavefunctions and charge density associated with the lowest levels of the topological superconducting nanowire spectrum are presented in Appendix \ref{AppA}.


\section{Nanowire SNS junctions}
\label{sec1}
\begin{figure}
\centering
\includegraphics[width=.45\textwidth]{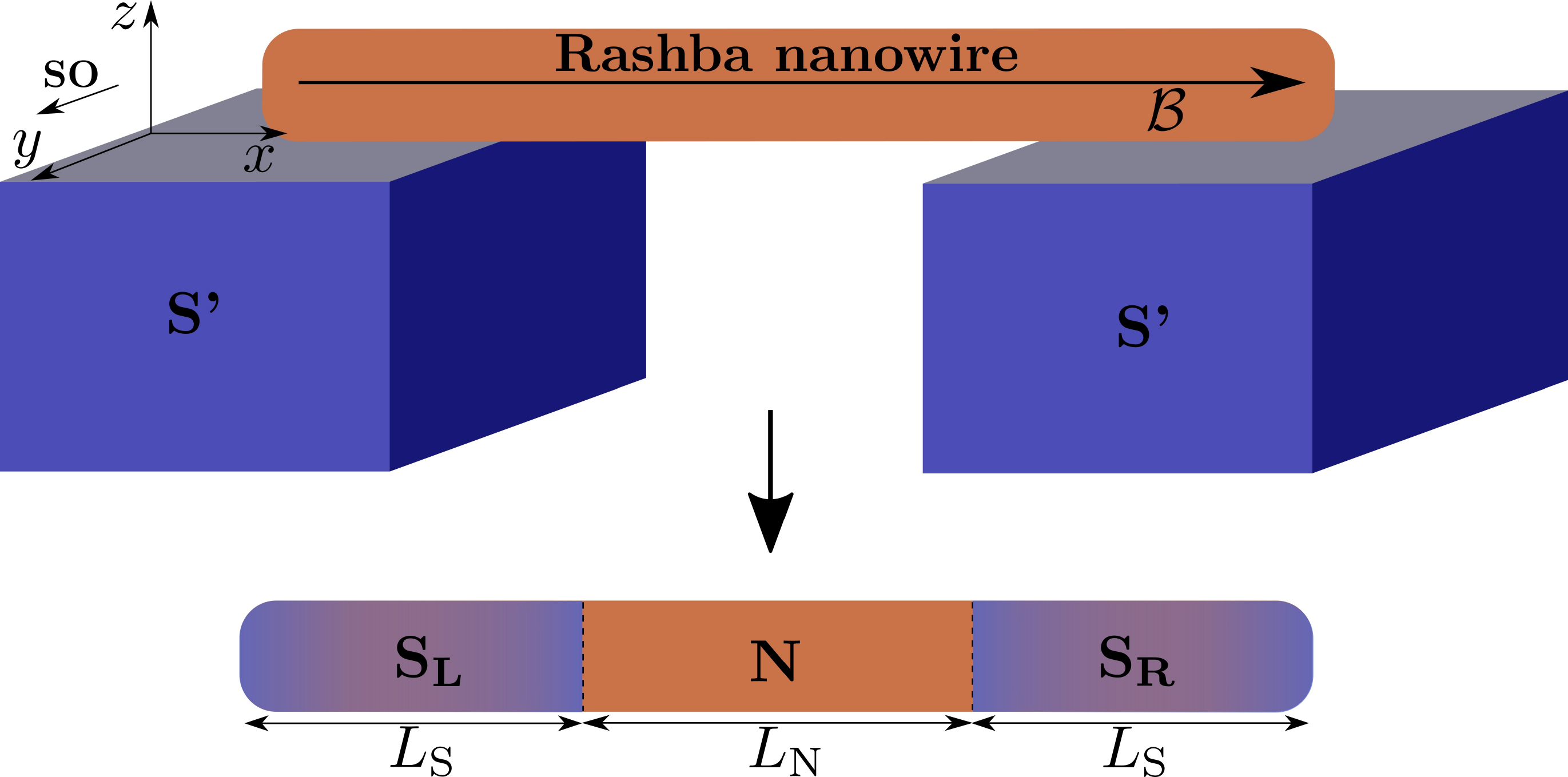} 
\caption{(Color online) Schematic of SNS junctions based on Rashba nanowires. Top: A nanowire with Rashba SOC of length $L=L_{\rm S}+L_{\rm N}+L_{\rm S}$ placed on top of two $s$-wave superconductors (S') with pairing potentials $\Delta_{S'}$ and subjected to an external magnetic field $\mathcal{B}$ (denoted by the black arrow). 
Superconducting correlations are induced into the nanowire via proximity effect. Bottom: Left and right regions of the nanowire become superconducting, denoted by $S_{L}$ and $S_{R}$, with induced pairing potentials 
$\Delta_{{\rm S}_{\rm L(R)}}<\Delta_{S'}$ and chemical potentials $\mu_{{\rm S}_{\rm L(R)}}$, while the central region remains in the normal state with $\Delta_{N}=0$ and chemical potential $\mu_{\rm N}$. This results in a Superconductor-Normal-Superconductor (SNS) junction.}
\label{fig1a}
\end{figure}

In this part, we concentrate on SNS junctions based on the proximitized nanowires that we discussed in the previous section. 
The basic geometry contains left (${\rm S}_{\rm L}$) and right (${\rm S}_{\rm R}$) superconducting regions of length $L_{\rm S}$ separated by a central normal (N) region of  length $L_{\rm N}$, as shown in Fig.\,\ref{fig1a}. 
The regions N and ${\rm S}_{\rm L(R)}$ are described by the tight-binding Hamiltonian $H_{0}$ given by Eq.\,(\ref{HamilTB}) with their respective chemical potentials, $\mu_{\rm N}$, $\mu_{{\rm S}_{\rm L(R)}}$. 

The Hamiltonian describing the SNS junction without superconductivity is then given by
\begin{equation}
\label{hsnszero}
h_{\rm SNS}=
\begin{pmatrix}
H_{\rm S_{L}}&H_{\rm S_{L}N}&0\\
H^{\dagger}_{\rm S_{L}N}&H_{\rm N}&H_{\rm NS_{R}}\\
0&H_{\rm NS_{R}}^{\dagger}&H_{\rm S_{R}}
\end{pmatrix}\,
\end{equation}
where $H_{\rm S_{i}}$ with $i=L/R$ and $H_{\rm N}$ are the Hamiltonians of the superconducting and normal regions, respectively. 
$H_{\rm S_{i}N}$ and $H_{\rm N S_{i}}$ are the Hamiltonians that couple  $S_{i}$ to the normal region $N$.
The elements of these coupling matrices are zero everywhere except for adjacent sites that lie at the interfaces of the S regions and of the N region.
In order to control the normal transmission $T_{\rm N}$ across these junctions,
 this coupling is parametrized by a hopping matrix $v_{0}=\uptau\,v$ between the sites that define the interfaces of the junction, where $\uptau\in[0,1]$.
The parameter $\uptau$ controls the normal transmission that ranges from fully transparent ($\uptau=1$) to tunnel ($\uptau\leq 0.6$), as discussed in Ref. \onlinecite{Cayao17b} for short junctions, being also valid for long junctions.  
The superconducting regions of the nanowire are characterized by chemical potential $\mu_{{\rm S}_{\rm L(R)}}$ and uniform superconducting pairing potential\cite{RevModPhys.51.101,Fer18} $\Delta_{{\rm S}_{\rm L}}=\bar{\Delta}\,{\rm e}^{-i\phi/2}$ and $\Delta_{{\rm S}_{\rm R}}=\bar{\Delta}\,{\rm e}^{i\phi/2}$, where $\Delta<\Delta_{\rm S'}$ and $\bar{\Delta}={\rm i}\sigma_{y}\Delta\,$. The central region of the nanowire is in the normal state without superconductivity, $\Delta_{\rm N}=0$, and with chemical potential $\mu_{\rm N}$.
Thus, the pairing potential matrix in the junction space reads
\begin{equation}
\label{deltasns}
\begin{split}
\Delta_{\rm SNS}&=
\begin{pmatrix}
\Delta_{\rm S_{L}}&0&0\\
0&\Delta_{\rm N}&0\\
0&0&\Delta_{\rm S_{R}}
\end{pmatrix}
=\begin{pmatrix}
\Delta_{0,S}\,{\rm e}^{{\rm i}\phi_{L}}&0&0\\
0&0&0\\
0&0&\Delta_{0,S}\,{\rm e}^{{\rm i}\phi_{R}}
\end{pmatrix}\,,
\end{split}
\end{equation} 
Next, we define the phase difference across the junction as $\phi_{\rm R}-\phi_{\rm L}=\phi$.

Therefore, the Hamiltonian for the full SNS junction reads in Nambu space \cite{PhysRevB.91.024514,Cayao17b}
\begin{equation}
\label{BdGS}
H_{\rm SNS}=
\begin{pmatrix}
h_{\rm SNS}&\Delta_{\rm SNS}\\
\Delta^{\dagger}_{\rm SNS}&-h_{\rm SNS}^{*}
\end{pmatrix}\,.
\end{equation}

In what follows, we discuss short ($L_{\rm N}\ll\xi$) and long ($L_{\rm N}\gg\xi$) SNS junctions, where $L_{\rm N}$ is the length of the normal region 
and $\xi=\hbar v_{F}/\Delta$ the superconducting coherence 
length.\cite{RevModPhys.51.101} The previous Hamiltonian is diagonalized numerically and in our calculations we consider realistic system parameters for InSb as described previously.


\subsection{Low-energy Andreev spectrum}
\label{sec11}
\begin{figure}
\centering
\includegraphics[width=.45\textwidth]{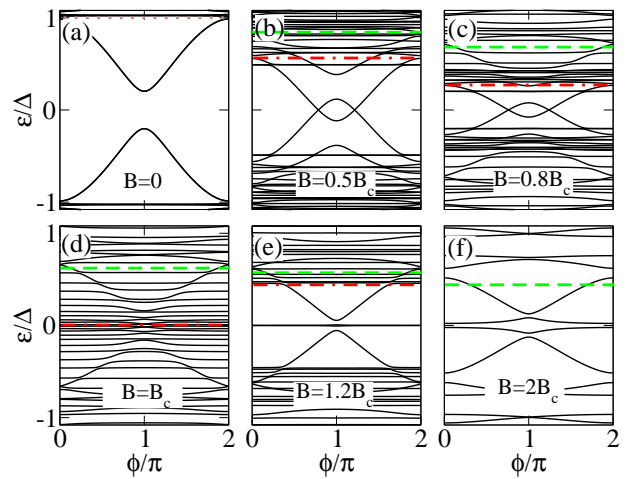} 
\caption{(Color online) Low-energy Andreev spectrum as a function of the superconducting phase  difference $\phi$ in a short SNS junction with $L_{\rm N}=20$\,nm and $L_{\rm S}=2000$\,nm. Different panels show the evolution with the Zeeman field: trivial phase for $B<B_{\rm c}$, topological transition at $B=B_{\rm c}$, and in the topological phase for $B>B_{\rm c}$. The energy spectrum exhibits the two different gaps which appear in the system for finite Zeeman field (marked by red and green dashed horizontal lines). Notice that after the gap inversion at $B=B_{\rm c}$, two MBSs emerge at the ends of the junction as almost dispersionless levels (outer MBSs), while additionally two MBSs appear at $\phi=\pi$ (inner MBSs). Parameters: $\alpha_{R}=20$\,meVnm, $\mu_{\rm N}=\mu_{\rm S}=0.5$\,meV and $\Delta=0.25$\,meV.}
\label{fig1}
\end{figure}

\begin{figure}
\centering
\includegraphics[width=.45\textwidth]{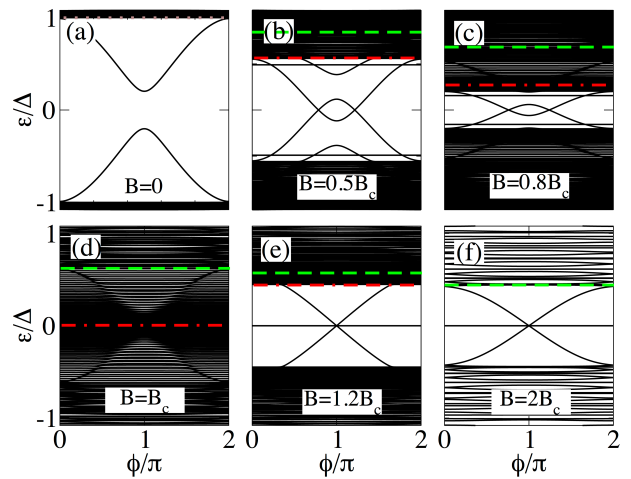} 
\caption{(Color online) Low-energy Andreev spectrum as a function of the superconducting phase difference in a short SNS junction with $L_{\rm N}=20$\,nm and $L_{\rm S}=10000$\,nm. Notice that in this case, the emergent outer MBSs are dispersionless with $\phi$, while the inner ones touch zero at $\phi=\pi$ acquiring Majorana character. 
Parameters: $\alpha_{R}=20$\,meVnm, $\mu_{\rm N}=\mu_{\rm S}=0.5$\,meV and $\Delta=0.25$\,meV.}
\label{fig1Ap}
\end{figure}

\begin{figure}
\centering
\includegraphics[width=.45\textwidth]{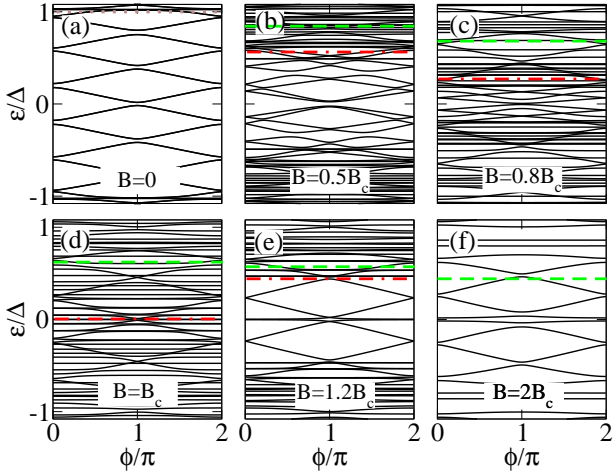} 
\caption{(Color online) Same as in Fig.\,\ref{fig1} for a long SNS junction with $L_{\rm N}=2000$\,nm and $L_{\rm S}=2000$\,nm. Notice that the four lowest states for $B>B_{\rm c}$ coexist with additional levels within $\Delta$.}
\label{fig2}
\end{figure}

\begin{figure}
\centering
\includegraphics[width=.45\textwidth]{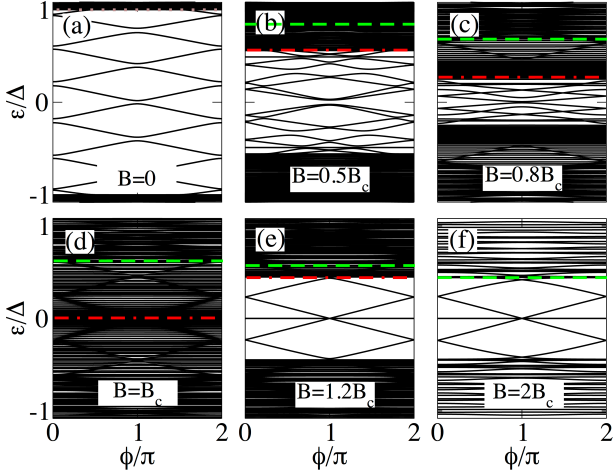} 
\caption{(Color online)(Color online) Same as in Fig.\,\ref{fig1Ap} for a long SNS junction with $L_{\rm S}=10000$\,nm. Note that in this case, the outer MBSs lie at zero energy and the inner ones reach zero at $\phi=\pi$ acquiring Majorana character. Notice that the four lowest levels coexist with additional levels which arise because $L_{\rm N}$ is longer.}
\label{fig2Ap}
\end{figure}

Now, we are in position to investigate the low-energy Andreev spectrum in short ($L_{\rm N}\ll \xi$) and long ($L_{\rm N}\gg\xi$) SNS junctions. In particular, we discuss the formation of Andreev bound states and their evolution from the trivial ($B<B_{\rm c}$) into the topological phases ($B>B_{\rm c}$). For this purpose we focus on the phase and Zeeman dependent low-energy spectrum in short and long junctions, presented in Figs.\,\ref{fig1} and \ref{fig2}  for $L_{\rm S}\leq2\xi_{\rm M}$. 
For completeness we also present the case of $L_{\rm S}\gg2\xi_{\rm M}$ in Figs.\,\ref{fig1Ap} and \ref{fig2Ap}.

We first discuss short junctions with $L_{\rm S}\leq2\xi_{\rm M}$. In this regime at $B=0$ two degenerate ABSs appear within $\Delta$ as solutions to the BdG equations described by Eq.\,(\ref{BdGS}), see Fig.\,\ref{fig1}(a). Note that even this non-topological case is anomalous as 
the ABS energies do not reach zero at $\phi=\pi$, unlike predicted by the standard theory for a transparent channel  within the Andreev approximation $\mu_{\rm S}\gg \Delta$.\cite{Beenakker:92} The dense amount of levels above $|\varepsilon_{p}|>\Delta$ represents the quasi-continuum of states, which consists of a discrete set of levels due to the finite length of the N and S regions. 
Moreover, it is worth to point out that the detachment (space between the ABSs and quasi-continuum) of the quasi-continuum at $\phi=0$ and $2\pi$  is not zero, something that strongly depends on the finite length of the S regions (see Fig.\,\ref{fig1Ap}).

For a non-zero Zeeman field, Fig.\,\ref{fig1}(b,c), the ABSs split and the two different gaps $\Delta_{1}$ and $\Delta_{2}$, discussed in Sec.\,\ref{sec0}, emerge indicated by  dash-dot red and dashed green lines, respectively.
By increasing the Zeeman field, the low momentum gap $\Delta_{1}$ gets reduced (dash-dot red line), as expected, while the gap $\Delta_{2}$ (dashed green line) remains finite although it gets slightly reduced as shown in Fig.\,\ref{fig1}(b,c). For stronger, but unrealistic values of, SOC we have checked that $\Delta_{2}$ is indeed constant. Notice that the two lowest levels in this regime, within $\Delta_{1}$ (dash-dot red line), develop a loop with two crossings that are independent of the S regions length but exhibit a strong dependence on the SOC, Zeeman field and chemical potential. Indeed, we have checked that they appear due to the interplay of SOC and Zeeman field and disappear when $\mu$ acquires very large values, namely, in the Andreev approximation.

At $B=B_{\rm c}$, the energy spectrum exhibits the closing of the low momentum gap $\Delta_{1}$, as indicated by red dash-dot line in Fig.\,\ref{fig1}(d). This signals the topological phase transition, since two gapped topologically different phases can only be connected through a gap closing.
By further increasing the Zeeman field, Figs.\,\ref{fig1}(e,f), $B>B_{\rm c}$, the inner gap $\Delta_{1}$ acquires a finite value again; this reopening of the gap $\Delta_{1}$ indicates that the system enters into the topological phase and the superconducting regions denoted by ${\rm S}_{\rm L(R)}$ become topological, while the $N$ region remains in the normal state.
Thus, MBSs are expected to appear for $B>B_{\rm c}$ at the ends of the two topological superconducting sectors, since they define interfaces between topologically different regions. 

This is what we indeed observe for $B>B_{\rm c}$ in Fig.\,\ref{fig1}(e,f), where the low energy spectrum has Majorana properties. In fact,
for $B>B_{\rm c}$, the topological phase is characterised by the emergence of two (almost) dispersionless levels with $\phi$, which represent the outer MBSs $\gamma_{1,4}$ formed at the ends of the junction. Additionally, the next two energy levels strongly depend on $\phi$ and tend towards zero at $\phi=\pi$, representing the inner MBSs $\gamma_{2,3}$ formed inside the junction. For sufficiently strong fields, $B=2B_{\rm c}$, the lowest gap is $\Delta_{2}$ indicated by green dashed line, which in principle bounds the MBSs. The quasi-continuum in this case corresponds to the discrete spectrum above and below $\Delta_{2}$, where $\Delta_{2}$ is the high momentum gap marked by green dashed horizontal line in  Fig.\,\ref{fig1}(e,f).
The four MBSs develop a large splitting around $\phi=\pi$ which arises 
when the MBSs wave-functions have a finite spatial overlap $L_{\rm S}\leq2\xi_{\rm M}$.
 Since the avoided crossing between the dispersionless levels (belonging to $\gamma_{1,4}$) and the dispersive levels 
(belonging to $\gamma_{2,3}$) around $\phi=\pi$ depends on the overlap between MBSs on each topological segment, it can be used to quantify the degree of Majorana non-locality.\footnote{A variant of this idea using quantum dot parity crossings has been discussed in Refs.\,\onlinecite{Prada:PRB17}, \onlinecite{Marcus17,PhysRevB.96.195430}.}
 This can be explicitly checked by considering SNS junctions with longer superconducting regions, where the condition $L_{\rm S}\gg2\xi_{\rm M}$ is fulfilled such that the energy splitting at $\phi=\pi$ is reduced.
As an example, we present in Fig.\,\ref{fig1Ap} the energy levels as a function of the phase difference for $L_{\rm S}\gg2\xi_{\rm M}$, where the low-energy spectrum undergoes some important changes.
First, we notice in Fig.\,\ref{fig1Ap} that the energy spectrum at $B=0$ for $|\varepsilon_{p}|>\Delta$, exhibits a visible denser spectrum than in Fig.\,\ref{fig1} signalling the quasi-continuum of states. 
Notice that in the topological phase, $B>B_{\rm c}$, the lowest two energy levels, associated to the outer MBSs, are insensitive to $\phi$ remaining at zero energy. Thus, they can be considered as truly zero modes.
On the other hand,  the inner MBSs are truly bound within $\Delta_{2}$, unlike Fig.\,\ref{fig1}, and  for $\phi=0$ and $\phi=2\pi$ they merge with the quasi-continuum at $\pm\Delta$. Thus, an increase in the length of the superconducting regions favours the reduction of the detachment between the discrete spectrum and the quasi-continuum at $0$ and $2\pi$, as it should be for a ballistic junction.\cite{Kwon:EPJB03, Fu:PRB09} Moreover, the energy splitting at $\phi=\pi$ is considerable reduced, even negligible. However, it will be always non-zero, though not visible to the naked eye, due to the finite length and thus due to the presence of the outer MBSs.  

The low-energy spectrum as a function of the superconducting phase difference for different values of the Zeeman field in long SNS junctions ($L_{\rm N}\gg\xi$) is presented in Fig.\,\ref{fig2}  for $L_{\rm S}\leq2\xi_{\rm M}$. We additional show in  Fig.\,\ref{fig2Ap} the case for $L_{\rm S}\gg2\xi_{\rm M}$.
\begin{center}
 \begin{figure}
\centering
\includegraphics[width=.45\textwidth]{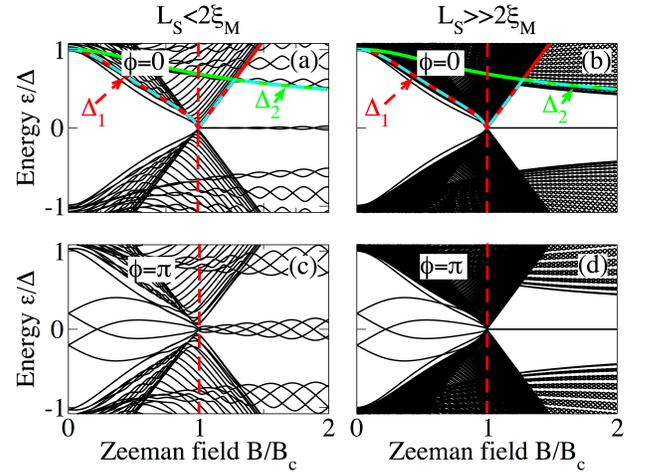} 
\caption{(Color online) Low-energy Andreev spectrum as a function of the Zeeman field in a short SNS junction at (a,b) $\phi=0$ and (c,d) $\phi=\pi$ with (a,c) $L_{\rm S}=2000$\,nm and (b,d) $L_{\rm S}=10000$\,nm. Notice that the low-energy spectrum traces the gap closing and reopening indicated by the solid red curve that corresponds to $\Delta_{1}$, while for $B>B_{\rm c}$ we observe the emergence of (a)  two at $\phi=0$ and (c) four MBSs at $\phi=\pi$  which oscillate around zero energy with $B$ due to $L_{\rm S}\leq2\xi_{\rm M}$ within a minigap defined by $\Delta_{2}$ (solid green curve). 
(b,d) For  $L_{\rm S}\gg2\xi_{\rm M}$ the MBSs are truly zero modes. Parameters: $L_{\rm N}=20$\,nm, $\alpha_{R}=20$\,meVnm, $\mu=0.5$\,meV and $\Delta=0.25$\,meV.}
\label{SNSa}
\end{figure}
\end{center}
As expected, long junctions contain more levels within the energy gap $\Delta$, see Figs.\,\ref{fig2}(a) and \ref{fig2Ap}(a), as compared with short junctions. As we shall discuss, this eventually affects the signatures of Majorana origin for $B>B_{\rm c}$, namely, the ones corresponding the the lowest four levels, in the supercurrents.

 \begin{figure}
\centering
\includegraphics[width=.45\textwidth]{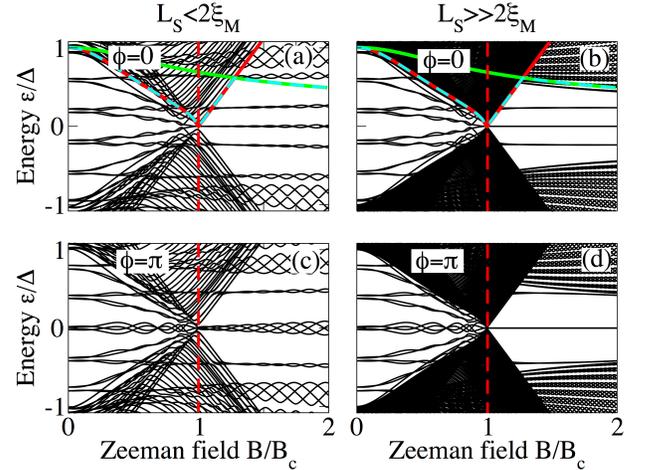} 
\caption{(Color online) Same as in Fig.\,\ref{SNSa} but for a long junction. 
Parameters: $L_{\rm N}=2000$\,nm, $\alpha_{R}=20$\,meVnm, $\mu=0.5$\,meV and $\Delta=0.25$\,meV.}
\label{SNSc}
\end{figure}

The above discussion can be clarified by considering the dependence of the low-energy spectrum on the Zeeman field at fixed phase difference $\phi=0$ and $\phi=\pi$. This is shown in Fig.\,\ref{SNSa} (short junction limit) and Fig.\ref{SNSc} (long junction limit) for  $L_{\rm S}\leq2\xi_{\rm M}$ (a,c)  and $L_{\rm S}\gg2\xi_{\rm M}$ (b,d), and for intermediate junctions in Fig.\,\ref{SNScx} of the Appendix. In panels (a,b) the gaps $\Delta_{1}$, $\Delta_{2}$ and ${\rm min}(\Delta_{1},\Delta_{2})$ in solid red, solid green and dashed cyan is also plotted to visualize the gap closing and reopening discussed in previous section. In all cases, it is clear that MBS smoothly evolve from the lowest ABS either following the closing of the induced gap, $\Delta_{1}$ indicated by solid red  curve, at $\phi=0$ or evolving from the lowest detached levels at $\phi=\pi$. The latter first cross zero energy, owing to Zeeman splitting, and eventually become four low-energy levels  oscillating out of phase around zero energy (Fig.\,\ref{SNSa}(c)).
The emergence of these oscillatory low-energy levels, separated by a minigap ($\Delta_{2}$ indicated by solid green curve) from the quasi-continuum,  characterizes the topological phase of the SNS junction. As expected, the oscillations get completely reduced for $L_{\rm S}\gg2\xi_{\rm M}$ and the four levels at $\phi=\pi$ become degenerate at zero energy, see Fig.\,\ref{SNSa}(b,d).

An increase in the length of the normal section results in an increase of the amount of subgap levels as observed in Figs.\,\ref{SNScx} and \,\ref{SNSc}. In both cases, in the topological phase, $B>B_{\rm c}$, these additional levels reduce the minigap with respect to short junctions and also slightly reduce the amplitude of the oscillations of the energy levels around zero as seen in Figs.\,\ref{SNScx}(a,b) and \ref{SNSc}(a,b). Also, the minigaps for $\phi=0$ and $\phi=\pi$ are different, in contrast to short junctions. In fact, the minigap at  $\phi=0$ is almost half of the value at $\phi=\pi$ for the chosen parameters. This can be understood from the phase dispersion of the long junction ABS spectra such as the ones in Figs. \ref{fig2} and \ref{fig2Ap}. For longer N regions this difference can be even larger.

From the above discussion it is clear that 
the energy spectrum of SNS nanowire junctions offers the possibility to directly monitor the ABSs which trace the gap inversion  and eventually evolve into MBSs.

\subsection{Supercurrents}
\label{sec12}
After having established in detail the energy spectrum in short and long SNS junctions, we now turn our attention to the corresponding phase-dependent supercurrents. They can be calculated directly from the discrete Andreev spectrum 
$\varepsilon_{p}$ as \cite{Beenakker:92,zagoskin,Cayao17b}
\begin{equation}
\label{josehpcurrent}
I(\phi)=-\frac{e}{\hbar}\sum_{p>0}{\rm tanh}\Big(\frac{\varepsilon_{p}}{2\kappa_{B}^{}T} \Big)\frac{d\varepsilon_{p}}{d\phi}\,,
\end{equation}
where $\kappa_{B}$ is the Boltzmann constant, $T$ the temperature and the summation is performed over positive eigenvalues $\varepsilon_{p}$. 
By construction, our junctions have finite length which implies that Eq.\,(\ref{josehpcurrent}) exactly includes the discrete quasi-continuum contribution.

\begin{figure}
\centering
\includegraphics[width=.45\textwidth]{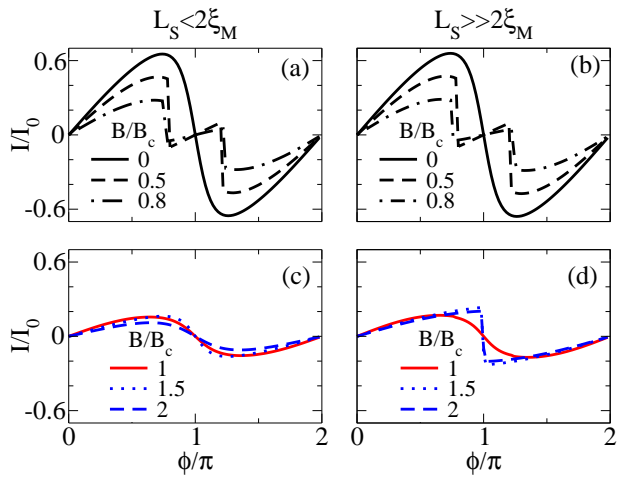} 
\caption[Josephson current in a short SNS junction]{(Color online) Supercurrent as a function of the superconducting phase difference in a short SNS junction, 
$I(\phi)$, for  (left column)  $L_{\rm S}=2000\,\text{nm}\leq2\xi_{\rm M}$ and (right column) $L_{\rm S}=10000\,\text{nm}\gg2\xi_{\rm M}$. Top row shows the Josephson current in the trivial phase for different values of 
the Zeeman field, $B<B_{\rm c}$, while bottom row corresponds to different values of the Zeeman field in the topological phase, $B\geq B_{\rm c}$. 
Notice the sawtooth feature at $\phi=\pi$ for $L_{\rm S}\gg2\xi_{\rm M}$.
 Parameters:  $\alpha_{R}=20$\,meVnm, $\mu=0.5$\,meV, $\Delta=0.25$meV and $I_{0}=e\Delta/\hbar$.}
\label{Fig8}
\end{figure}

\begin{figure}
\centering
\includegraphics[width=.45\textwidth]{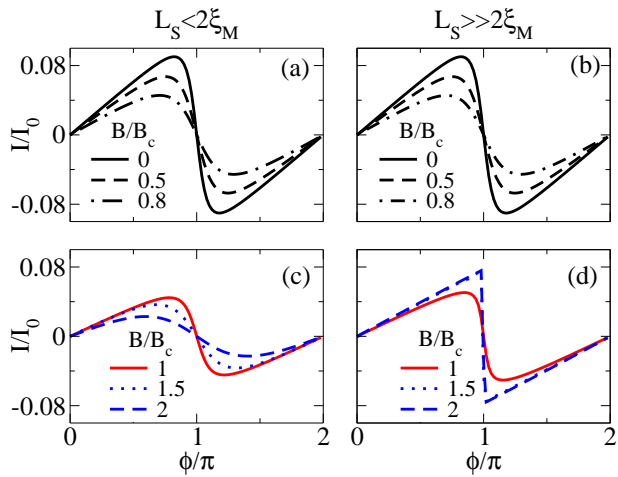} 
\caption[Josephson current in a long SNS junction]{(Color online) Same as in Fig.\,\ref{Fig8} for a long SNS junction, $L_{\rm N}=2000$\,nm. Note that in this case the magnitude of the supercurrent is reduced, an effect caused by the length of the normal section.}
\label{Fig9}
\end{figure}

In Figs.\,\ref{Fig8} and \ref{Fig9} we present supercurrents as a function of the superconducting phase difference $I(\phi)$ for different values of the Zeeman field in short and long SNS junctions, respectively.  Panels (a,c) correspond to $L_{\rm S}\leq2\xi_{\rm M}$, while (b,d) to $L_{\rm S}\gg2\xi_{\rm M}$.

Firstly, we discuss the short junction regime, presented in Fig.\,\ref{Fig8}.  At $B=0$ the supercurrent $I(\phi)$ has a sine-like dependence on $\phi$, with a relative fast change of sign around $\phi=\pi$ that is determined by the derivative of the lowest energy spectrum profile around $\phi=\pi$. This result is different from usual ballistic full transparent supercurrents in trivial SNS junctions,\cite{Beenakker:92} where the supercurrent is proportional to ${\rm sin}(\phi/2)$ being maximum at $\phi=\pi$. This difference from the standard ballistic limit can be ascribed to deviations from the ideal Andreev approximation, see also the discussion of Fig.\,\ref{fig1}(a), owing to the relatively low chemical potential needed to reach the helical limit and, eventually, the topological regime as Zeeman increases. At finite values of the Zeeman field $B$, but still in the trivial phase $B<B_{\rm c}$ (dashed and dash-dot curves), $I(\phi)$ undergoes some changes. First, the maximum value of $I(\phi)$ is reduced due to the reduction of the induced gap that is caused by the Zeeman effect. Second, $I(\phi)$ develops a zig-zag  profile (just before and after $\phi=\pi$) signalling a 0-$\pi$ transition in the supercurrent. This transition arises from the zero-energy crossings in the low-energy spectrum, see Fig.\,\ref{fig1}(b,c). As discussed in previously, these level crossings result from the combined action of both SOC and Zeeman field at low $\mu$, and introduce discontinuities in the derivatives with respect to $\phi$. 
Besides these features, all the supercurrent curves for $B<B_{\rm c}$, for both $L_{\rm S}\leq2\xi_{\rm M}$ and $L_{\rm S}\gg2\xi_{\rm M}$, exhibit a similar behaviour, see Fig.\,\ref{Fig8}. Interestingly, the system is gapless at the topological transition, $B=B_{\rm c}$, but the supercurrent remains finite, see red curve in Fig.\,\ref{Fig8}(c). 

For $B>B_{\rm c}$, the S regions of the SNS junction become topological and MBSs emerge at their ends, as described in the previous subsection.
Despite the presence of MBSs the supercurrent $I(\phi)$ remains $2\pi$-periodic,  namely, $I(\phi)=I(\phi+2\pi)$. This results from the fact that we sum up positive levels, only, as we deal with an equilibrium situation (since the supercurrent is a ground state property, transitions between the negative and positive energies around are not allowed since there is an energy gap originated from Majorana overlaps). Strategies to detect the presence of MBSs beyond the equilibrium supercurrents described here include the ac Josephson effect, noise measurements, switching-current measurements, microwave spectroscopy, dynamical susceptibility measurements, etc.\cite{Badiane:PRL11,San-Jose:11a,Pikulin:PRB12,PhysRevB.94.085409,PhysRevB.92.134508,Mirceacondmat17}

As the Zeeman field is further increased in the topological phase, $B>B_{\rm c}$, the  supercurrent tends to decrease due to the finite  Majorana overlaps when $L_{\rm S}\leq2\xi_{\rm M}$ , see dotted and dashed blue curves in Fig.\,\ref{Fig8}(d). On the other hand, as the length of S becomes larger such that $L_{\rm S}\gg2\xi_{\rm M}$ the overlap gets reduced, which is reflected in a clear \emph{sawtooth} profile at $\phi=\pi$, see dotted and dashed blue curves in Fig.\,\ref{Fig8}(d). This discontinuity in $I(\phi)$ depends on $L_{\rm S}$ and results from the profile of the lowest energy spectrum at $\phi=\pi$, as shown in Fig.\,\ref{fig1Ap}(d).  The sawtooth profile thus indicates the emergence of well-localized MBSs at the ends of S and represents one of our main findings. 

As discussed above, the supercurrent for $B<B_{\rm c}$, Figs.\,\ref{Fig8}(a,b), does not depend on $L_{\rm S}$. On the contrary, the supercurrents in the topological phase, Figs.\,\ref{Fig8}(c,d), do strongly depend on $L_{\rm S}$ owing to the emergence of MBSs. 

In Fig.\,\ref{Fig9} we present $I(\phi)$ for long junctions $L_{\rm N}\gg\xi$ at different values of the Zeeman field. Different panels correspond to (a,c) $L_{\rm S}\leq2\xi_{\rm M}$ and (b,d) $L_{\rm S}\gg2\xi_{\rm M}$.  
Notice that even though the maximum value of the current is reduced in this regime, the overall behaviour is very similar to the short junction regime for both $B<B_{\rm c}$ and $B>B_{\rm c}$.
The only important difference with respect to the short junction case is that $I(\phi)$ in the long junction regime does not exhibit the zig-zag profile due to 0-$\pi$ transitions.

As the system enters into the topological phase for $B>B_{\rm c}$ and $L_{\rm S}\leq2\xi_{\rm M}$, Fig.\,\ref{Fig9}(c), the maximum supercurrent decreases signalling the non-zero splitting at $\phi=\pi$ in the low energy spectrum. Deep in the topological phase, the supercurrent exhibits a slow (slower than in the trivial phase Fig.\,\ref{Fig9}(a)) sign change around $\phi=\pi$, and its dependence on $\phi$ tends to approach a sine function.
 However, for $L_{\rm S}\gg2\xi_{\rm M}$ shown in Fig.\,\ref{Fig9}(d), the supercurrent acquires an almost constant value as $B$ increases and  develops a clear sawtooth profile at $\phi=\pi$ due to the zero energy splitting in the low energy spectrum at $\phi=\pi$, similarly to the short junction case. It is worth to point out that, although the maximum supercurrent is slightly reduced, deep in the topological phase (dashed and dotted blue curves) its maximum value is approximately close to the maximum value in the trivial phase, which is different to what we found in the short junction case. This is a clear consequence of the emergence of additional levels within the induced gap  due to the increase of $L_{\rm N}$. These additional levels exhibit a strong dependence on the superconducting phase, very similar to the inner MBSs as one can see in Fig.\,\ref{fig2}(e,f). 

\begin{figure}
\centering
\includegraphics[width=.45\textwidth]{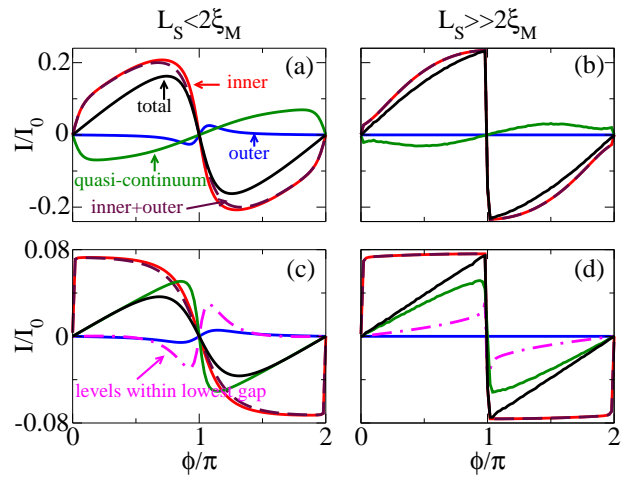} 
\caption[Josephson current in a long SNS junction]{(Color online) Supercurrent as a function of $\phi$ at $B=1.5B_{\rm c}$.  Contribution to supercurrent  in short  (a,b) and  long (c,d) junctions for (a,c)  $L_{\rm S}\leq2\xi_{\rm M}$ and (b,d) $L_{\rm S}\gg2\xi_{\rm M}$, respectively. Different curves in (a,b) correspond to individual contributions to $I(\phi)$ from outer, inner, outer$+$inner (levels within lowest induced gap $\Delta_{2}$), quasi-continuum (levels above the lowest gap $\Delta_{2}$) and total levels; while in (c,d) the additional magenta curve corresponds to all levels within $\Delta_{2}$. Notice that in long junctions the number of levels within $\Delta_{2}$ exceeds the number of MBSs. Notice that MBSs coexist with additional levels within $\Delta_{2}$.  
Parameters:  $\alpha_{R}=20$\,meVnm, $\mu=0.5$\,meV, $\Delta=0.25$meV and $I_{0}=e\Delta/\hbar$.}
\label{Fig10}
\end{figure}

In order to analyse the individual contribution of outer and inner MBSs with respect to the quasi-continuum we calculate  and identify supercurrents for such situations. This is presented in Fig.\,\ref{Fig10} for (a,b) short junctions and (c,d) for long junctions, for (a,c) $L_{\rm S}\leq2\xi_{\rm M}$ and (b,d) $L_{\rm S}\gg2\xi_{\rm M}$. Remembering that in this regime the lowest gap is the high momentum gap $\Delta_{2}$, and the levels outside this gap constitute the quasi-continuum.

Firstly, we discuss short junctions.
The supercurrent due to outer MBSs for $L_{\rm S}\leq2\xi_{\rm M}$  is finite only around $\phi=\pi$, exhibiting an odd dependence on $\phi$ around $\pi$; however, away from this point it is approximately zero and independent of $\phi$ (see blue curve in Fig.\,\ref{Fig10}(a)). When $L_{\rm S}\gg2\xi_{\rm M}$ the outer MBSs are very far apart and their contribution to $I(\phi)$ is zero, see blue curve in Fig.\,\ref{Fig10}(b).
On the other hand, the inner MBSs contribution to $I(\phi)$ is enormous and the outer part only slightly changes the shape of the maximum supercurrent when $L_{\rm S}\leq2\xi_{\rm M}$, while for $L_{\rm S}\gg2\xi_{\rm M}$ the outer MBSs do not contribute, as shown by dashed brown curve in Fig.\,\ref{Fig10}(a,b). Moreover, the inner contribution follows the roughly the same dependence on $\phi$ as the contribution of the whole energy spectrum shown in black curve in Fig.\,\ref{Fig10}(a,b). As described in the previous subsection, the quasi-continuum is considered to be the discrete energy spectrum above $|\varepsilon_{p}|>\Delta_{2}$.
The quasi-continuum contribution to $I(\phi)$ is finite and odd in $\phi$ around $\pi$, as shown by green curves in Fig.\,\ref{Fig10}(a,b). The quasi-continuum contribution  to the total supercurrent $I(\phi)$ far away from $\phi=\pi$ is appreciable mainly when the MBSs exhibit a finite energy splitting as seen in Fig.\,\ref{Fig10}(a). Interestingly, 
 the outer and in particular the inner MBSs (levels within $\Delta_{2}$) are the main source when such overlap is completely reduced and determine the profile of $I(\phi)$, as shown in Fig.\,\ref{Fig10}(b). 

In long junctions the situation is a bit different mainly because more levels emerge within $\Delta$ in the trivial phase. For $B>B_{\rm c}$ within $\Delta_{2}$ these additional levels coexist with the inner and outer MBSs, see Fig.\,\ref{Fig10}(c,d). The contribution of the outer MBSs to $I(\phi)$ exhibits roughly similar behaviour as for short junctions although smoother around $\phi=\pi$ , shown by blue curve in Fig.\,\ref{Fig10}(c,d). The inner MBSs, however, have a strong dependence on $\phi$ and develop their maximum value close to $\phi=2\pi n$ with $n=0,1,\cdots$ (see red curve), unlike short junctions. The outer MBSs almost do not affect the overall shape of the $I(\phi)$ curve (see dashed brown curve). Since a long junction host more levels, we also show in dash-dot magenta curve the contribution of all the levels within $\Delta_{2}$, including also the four MBSs. Notice that such contribution is considerable large close to $\phi=\pi$, only, with a minimum and maximum value before and after $\phi=\pi$ for $L_{\rm S}\leq2\xi_{\rm M}$, respectively. This is indeed the reason why the supercurrent gets reduced as $B$ increases in the topological phase for $L_{\rm S}\leq2\xi_{\rm M}$, see dotted and dashed blue curves in Fig.\,\ref{Fig9}(c).
On the other hand for  $L_{\rm S}\gg2\xi_{\rm M}$ the contribution of all the levels within $\Delta_{2}$ exhibits a sawtooth profile at $\phi=\pi$, which, instead of reducing the quasi-continuum contribution (green curve), increase the maximum value of $I(\phi)$ at $\phi=\pi$ resulting in the solid black curve. Importantly, unlike short junctions, in long junctions the quasi-continuum modifies the $I(\phi)$ around $\phi=\pi$. Thus, a zero temperature current-phase measurement in a SNS junction setup could indeed reveal the presence of MBSs by  observing the reduction of the maximum supercurrent. In particular, well localized MBSs are revealed in the sawtooth profile of $I(\phi)$  at $\phi=\pi$.
In what follows we analyse the effect of temperature, variation of normal transmission and random disorder on the sawtooth profile at $\phi=\pi$ of the supercurrent.

\begin{figure}
\centering
\includegraphics[width=.45\textwidth]{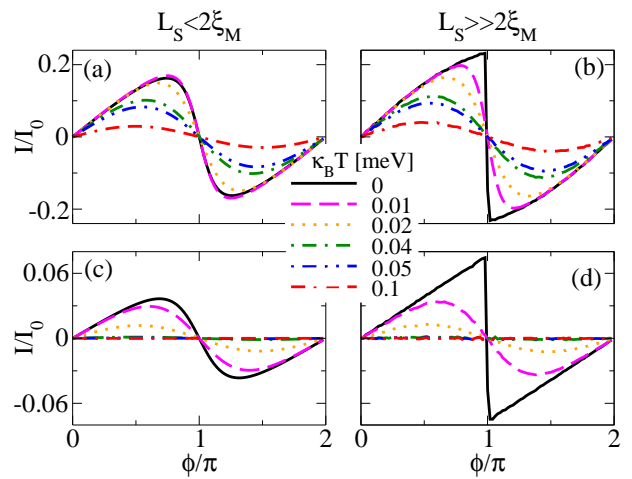} 
\caption[Josephson current in a long SNS junction]{(Color online) Finite temperature effect on the supercurrent  $I(\phi)$ in a short (a,b) and long (c,d) SNS junction at $B=1.5B_{\rm c}$ for  (left)  $L_{\rm S}=2000\,\text{nm}\leq2\xi_{\rm M}$ and (right) $L_{\rm S}=10000\,\text{nm}\gg2\xi_{\rm M}$. Different curves corresponds to different values of $\kappa_{B}T$. Notice that the \emph{sawtooth} profile smooths out at finite temperature.
 Parameters: $L_{\rm N}=20$\,nm for short and $L_{\rm N}=2000$\,nm for long junctions, $\alpha_{R}=20$\,meVnm, $\mu=0.5$\,meV, $\Delta=0.25$meV and $I_{0}=e\Delta/\hbar$.}
\label{IT}
\end{figure}
{\bf Temperature effects:}
In this part, we analyse the effect of temperature on supercurrents in the topological phase. 
In Fig.\,\ref{IT} we present the supercurrent as a function of the superconducting phase difference $I(\phi)$ in the topological phase $B=1.5B_{\rm c}$ at different temperature values for (a) $L_{\rm S}\leq2\xi_{\rm M}$ and (b) $L_{\rm S}\gg2\xi_{\rm M}$.
At zero temperature for $L_{\rm S}\leq2\xi_{\rm M}$, shown by black solid curve in (a), the dependence of the supercurrent on $\phi$ approximately corresponds to a sine-like function of $\phi$, as pointed out before and  a small increase in temperature $\kappa_{\rm B}T=0.01$\,meV (magenta dashed curve) slightly modifies the profile of the maximum supercurrent. However, for $L_{\rm S}\gg2\xi_{\rm M}$ in (b), the same temperature (dashed curve) value has a detrimental effect on the sawtooth profile of $I(\phi)$ at $\phi=\pi$ that reduces its maximum value and smooths out due to thermal population of ABSs. We have checked that smaller temperature values than the ones presented in Fig.\,\ref{IT} indeed still smooths out such sawtooth profile but the fast sign change around $\phi=\pi$ is still visible: this effect remains as long as  $\kappa_{\rm B}T\ll\Delta$. As  temperature increases, $I(\phi)$ in (a,b) smoothly acquires a true sine function of $\phi$, as seen in Fig.\,\ref{IT}(a). Although the sawtooth profile might be hard to observe in real experiments, the maximum value of $I(\phi)$, which is finite in the topological phase and almost halved with respect to the trivial phase in short junctions,\cite{Cayao17b} still provide a tool for distinguishing it from $I(\phi)$ in trivial junctions.

\begin{figure}
\centering
\includegraphics[width=.45\textwidth]{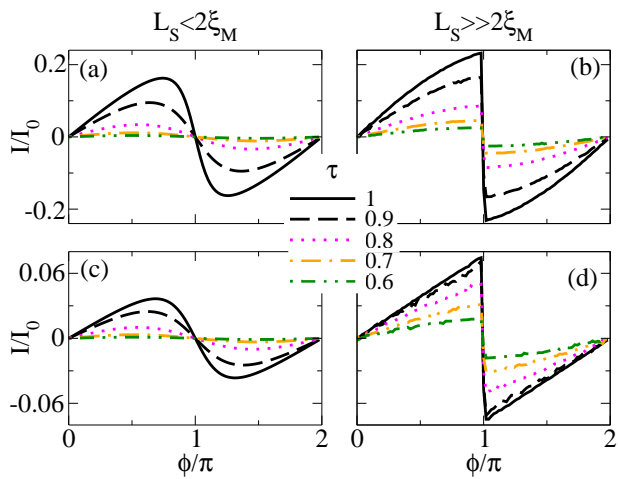} 
\caption[Josephson current in a long SNS junction]{(Color online) Effect of normal transmission through coupling parameter $\uptau$ on the supercurrent  $I(\phi)$ in a short (a,b) and long (c,d) SNS junction at $B=1.5B_{\rm c}$ for  (left)  $L_{\rm S}\ll2\xi_{\rm M}$, $L_{\rm S}=2000$\,nm, and (right) $L_{\rm S}\gg2\xi_{\rm M}$, $L_{\rm S}=10000$\,nm. Notice that, although by 
decreasing $\uptau$  the magnitude of the supercurrent at $\phi=\pi$ decreases, the \emph{sawtooth} profile is preserved.
Parameters: $L_{\rm N}=20$\,nm for short and $L_{\rm N}=2000$\,nm for long junctions, $\alpha_{R}=20$\,meVnm, $\mu=0.5$\,meV, $\Delta=0.25$meV and $I_{0}=e\Delta/\hbar$.}
\label{Itau}
\end{figure}
{\bf Normal transmission effects:}
The assumption of perfect coupling between N and S regions in previous calculations is indeed a good approximation due to the enormous advance in fabrication of hybrid systems. However, it is also relevant to study whether the sawtooth profile of $I(\phi)$ is preserved or not when the normal transmission $T_{\rm N}$, controlled by $\uptau$, is varied. 

Fig.\,\ref{Itau} shows the supercurrent $I(\phi)$ in short junctions at $B=1.5B_{\rm c}$ for different values of $\uptau$ for (a) $L_{\rm S}\leq2\xi_{\rm M}$  and (b) $ L_{\rm S}\gg2\xi_{\rm M}$. By reducing $\uptau$ the supercurrent $I(\phi)$ is also reduced in both cases (a,b). However, for (a) $L_{\rm S}\leq2\xi_{\rm M}$ it experiments a transition from a sudden sign change around $\phi=\pi$ to a true sine function of $\phi$ when reducing $\uptau$, very similar to the effect of temperature discussed above. Notice that in the tunnel regime, $\uptau=0.6$, $I(\phi)$ is considerable reduced and approximately zero.
On the other hand, for $L_{\rm S}\gg2\xi_{\rm M}$ the sawtooth profile at $\phi=\pi$ is preserved and robust when $\uptau$ is reduced from the full transparent to the tunnel regime, as seen in Fig.\,\ref{Itau}(b). Quite remarkably, in the tunneling regime, $I(\phi)$ is finite away from $n\pi$ for $n=0,1,\cdots$, unlike $L_{\rm S}\leq2\xi_{\rm M}$ presented in (a).  The finite value of the supercurrent indeed could serve as another indicator of the non-trivial topology and thus of the emergence of MBSs in the junction.

{\bf Disorder effects:}
Now we analyse the sawtooth profile of $I(\phi)$ for $B>B_{\rm c}$ in presence of disorder. In particular, disorder is introduced as a random onsite potential $V_{i}$ in the tight-binding Hamiltonian given by Eq.\,(\ref{HamilTB}), where  the values of $V_{i}$ lie within $[-w,w]$, being $w$ the disorder strength. This kind of disorder introduces random fluctuations in the chemical potential of the system $\mu$ and therefore reasonable values of $w$ do not include $w\gg\mu$.

\begin{figure}
\centering
\includegraphics[width=.45\textwidth]{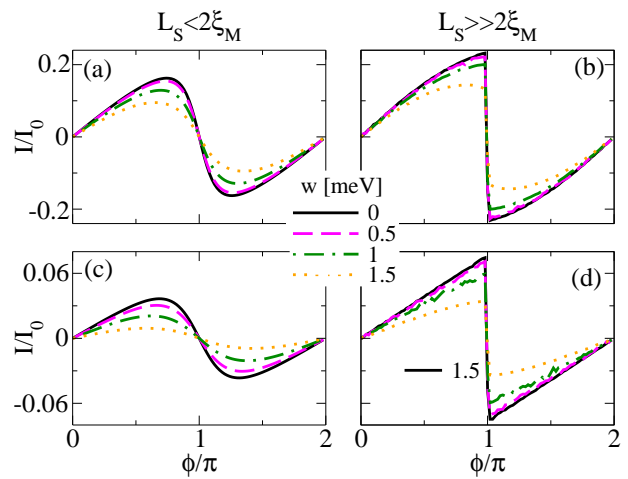} 
\caption[Josephson current in a long SNS junction]{(Color online) Effect of random on-site scalar disorder on the supercurrent  $I(\phi)$ in a short (a,b) and long (c,d) SNS junction at $B=1.5B_{\rm c}$ for  (left)  $L_{\rm S}\ll2\xi_{\rm M}$, $L_{\rm S}=2000$\,nm, and (right) $L_{\rm S}\gg2\xi_{\rm M}$, $L_{\rm S}=10000$\,nm. Each curve corresponds to 20 realizations of disorder, where $w$ is the disorder strength. Notice that for small $w$ of the order of the chemical potential, the \emph{sawtooth} profile at  $\phi=\pi$ is preserved (see right panel).
Parameters: $L_{\rm N}=20$\,nm for short and $L_{\rm N}=2000$\,nm for long junctions,  $\alpha_{R}=20$\,meVnm, $\mu=0.5$\,meV, $\Delta=0.25$meV and $I_{0}=e\Delta/\hbar$.}
\label{Idis}
\end{figure}

In Fig.\,\ref{Idis}(a,b) we present $I(\phi)$ in short junctions at $B=1.5B_{\rm c}$ for 20 disorder realizations and different values of the disorder strength $w$. Disorder of the order of the chemical potential $\mu$ has little effect on $I(\phi)$ as shown by dashed curves in Fig.\,\ref{Idis}(a,b), where the behaviour of $I(\phi)$ is approximately the same as without disorder. This reflects the robustness of the topological phase, and thus of MBSs, against fluctuations in the chemical potential.\cite{PhysRevB.95.184511,PhysRevB.96.165415} Stronger disorder (dotted and dash-dot curves) reduce the maximum value of $I(\phi)$, though its general behaviour is preserved. In fact, observe that the sawtooth profile at $\phi=\pi$ in (b) is robust against moderate disorder $w$. We have checked that these conclusions remain even when we consider disorder of the order of $5\mu$ (not shown).

\section{Conclusions}
\label{concl}
In this numerical work we have performed a detailed investigation of the low-energy spectrum and supercurrents in short ($L_{\rm N}\ll\xi$) and long  ($L_{\rm N}\gg\xi$)  SNS junctions based on nanowires with Rashba SOC and in the presence of a Zeeman field.

In the first part, we have studied the evolution of the low-energy Andreev spectrum from the trivial phase  into the topological phase and the emergence of MBSs in short and long SNS junctions. 
We have shown that the topological phase is characterized by the emergence of four MBSs in the junction (two at the outer part of the junction and two at the inner part) with important consequences to the equilibrium supercurrent. 
In fact, the outer MBSs are almost dispersionless with  respect to superconducting phase $\phi$, while the inner ones  disperse and tend to reach zero at $\phi=\pi$. A finite energy splitting at $\phi=\pi$ occurs whenever the length of the superconducting nanowire regions, $L_{\rm S}$, is comparable or less than $2\xi_{\rm M}$. Although in principle such energy splitting can be reduced by making the S regions longer, we conclude that in a finite length system the current-phase curves are $2\pi$-periodic and the splitting always spoils the so-called $4\pi$-periodic fractional Josephson effect in an equilibrium situation.

In short junctions the four MBSs are truly bound within $\Delta$ only when  $L_{\rm S}\gg2\xi_{\rm M}$, while in long junctions the four MBSs coexist with additional levels which profoundly affects phase-biased transport. 
As the Zeeman field increases in the trivial phase $B<B_{\rm c}$, the supercurrent  $I(\phi)$ gets reduced due to the reduction of the induced gap. In this case, the supercurrents $I(\phi)$ are independent of the superconducting regions length $L_{\rm S}$, an effect preserved in both short and long junctions. 

In short junctions in the topological phase with $B>B_{\rm c}$ the contribution of the four MBSs levels within the gap determines the shape of the current-phase curve $I(\phi)$ with only little contribution from the quasi-continuum. For $L_{\rm S}<2\xi_{\rm M}$, the overlap of MBS wavefunctions at each S region is finite and the quasi-continuum contribution is appreciable and of the opposite sign than the bound states contribution, thus inducing a reduction of the  maximum supercurrent in the topological phase.
We found that for $L_{\rm S}\gg2\xi_{\rm M}$, when both the spatial overlap  between MBSs and the splitting at $\phi=\pi$ are negligible, the quasi-continuum contribution is very small and the supercurrent $I(\phi)$ is  dominated by the inner MBSs. Remarkably, we have demonstrated that the current-phase curve $I(\phi)$ develops a clear sawtooth profile at $\phi=\pi$, which is independent of the quasi-continuum contribution and represents a robust signature of MBSs.

In the case of long junctions we have found that the additional levels that emerge within the gap affect the individual MBSs contribution. Here, it is the combined contribution of the levels within the gap and the quasi-continuum that determine the full current-phase curve $I(\phi)$, unlike in short junctions. The maximum supercurrent in long junctions are reduced in comparison to short junctions, as expected. Our results also show that the maximum value of the supercurrent in the topological phase depends on $L_{\rm S}$, acquiring larger values for $L_{\rm S}\gg2\xi_{\rm M}$ than for $L_{\rm S}\leq2\xi_{\rm M}$.

Finally, we have analyzed the robustness of the characteristic  sawtooth profile in the topological phase against temperature, changes in transmission across the junction and random on-site scalar disorder. We found that a small finite temperature smooths it out due to thermal population of ABSs. We demonstrated that, although this might be a fragile indicator of MBSs, the fast sign change around $\phi=\pi$ could help to distinguish the emergence of MBSs from trivial ABSs. Remarkably, the sawtooth profile is preserved against changes in transmission: it is preserved even in the tunnelling regime.  And finally, we showed that reasonable fluctuations  in the chemical potential $\mu$ (up to five times $\mu$) do not affect the sawtooth profile of $I(\phi)$ at $\phi=\pi$.

Our main contribution are summarized as follows. 
In finite length short and long  SNS junctions  four MBSs emerge, two at the inner part of junction and two at the outer ends.
The unavoidable overlap of the four MBSs gives rise to a finite energy splitting at $\phi=\pi$, thus rendering the equilibrium Josephson effect 2$\pi$-periodic in both short and long junctions. 
Current-phase curves in short and long junctions exhibit a clear sawtooth profile when the energy splitting near $\phi=\pi$ is small, therefore signalling the presence of weakly overlapping MBSs.   
Remarkably, current-phase curves do not depend on $L_{\rm S}$ in the trivial phase for both short and long junctions, while they strongly depend on $L_{\rm S}$ in the topological phase. This effect is purely connected to the splitting of MBSs at $\phi=\pi$, indicating a unique feature due to the topological phase and therefore of the presence of MBSs in the junction.

\section{Acknowledgements}
J.C. thanks  O.~A.~Awoga, K.~Bj\"{o}rnson, M. Benito and S. Pradhan for motivating and helpful discussions.
J.C. and A.B.S acknowledge financial support from the Swedish Research Council (Vetenskapsr\aa det), the G\"{o}ran Gustafsson Foundation, the Swedish Foundation for Strategic Research (SSF), and the Knut and Alice Wallenberg Foundation through the Wallenberg Academy Fellows program.
We also acknowledge financial support from the Spanish Ministry of Economy and Competitiveness through Grant No. FIS2015-64654-P  (R. A.), FIS2016-80434-P (AEI/FEDER, EU) (E. P.) and the Ram\'{o}n y Cajal programme through grant No. RYC-2011-09345 (E. P). 
\bibliography{biblio}
\clearpage
\appendix
\section{Supplementary calculations}
\label{AppA}
In this part we provide further calculations in order to support our findings in the main text of our manuscript.

\subsection{Superconducting wire}
\label{scwiremodel}
From Eq.\,(\ref{SCNW}) we can also calculate the wavefunctions associated to the energy levels after diagonalization. According to the chosen basis of Eq.\,(\ref{SCNW}), it is obtained in the following form
\begin{equation}
\Psi(x)=
\begin{pmatrix}
u_{\uparrow,i},
u_{\downarrow,i},
v_{\uparrow,i},
v_{\downarrow,i}
\end{pmatrix}^{T}
\end{equation}
where $T$ denotes the transpose operation and $N_{\rm S}$ is the number of sites of the discretised superconducting nanowire.
Then, the BdG wavefunction amplitude is given by
\begin{equation}
\label{wavefunction}
|\Psi(x)|^{2}=|u_{\uparrow,i}|^{2}+|u_{\downarrow,i}|^{2}+|v_{\uparrow,i}|^{2}+|v_{\downarrow,i}|^{2}\,.
\end{equation}
Likewise, for the same price we can calculate the charge density as it has been shown to provide useful information regarding MBSs.\cite{Ben-Shach:PRB15} It
can be calculated by using the same information of $\Psi(x)$ and  reads
\begin{equation}
\label{chargedensity}
|\rho(x)|^{2}=|v_{\uparrow,i}|^{2}+|v_{\downarrow,i}|^{2}-|u_{\uparrow,i}|^{2}-|u_{\downarrow,i}|^{2}\,.
\end{equation}

\begin{figure}
\centering
\includegraphics[width=.45\textwidth]{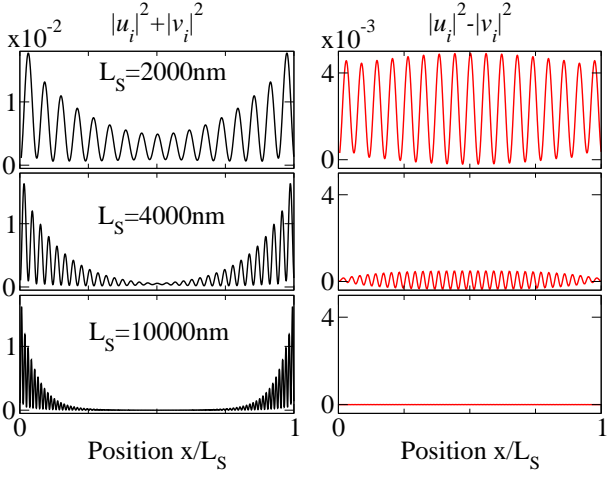} 
\caption{(Color online) Wavefunction amplitude $|\Psi(x)|^{2}$ and charge density $|\rho(x)|^{2}$, given by Eqs.\,(\ref{wavefunction}) and (\ref{chargedensity}), respectively, for different $L_{\rm S}$ corresponding to the two lowest levels (MBSs) in a topological superconducting nanowire. 
Parameters: $\alpha_{R}=20$\,meVnm, $\mu_{\rm N}=\mu_{\rm S}=0.5$\,meV, $\Delta=0.25$\,meV and $B=2B_{\rm c}$.}
\label{wavefunc}
\end{figure}

Thus, the wavefunction amplitude and charge density can be calculated after finding $\Psi(x)$. Now, we calculate them associated to the two lowest energy levels of the topological superconducting nanowire. 
This is presented in Fig.\,\ref{wavefunc} for different lengths of the wire in the topological phase, where left and right columns correspond to the wavefunction amplitude and charge density, respectively. 

Observe that for $L_{\rm S}=2000$\,nm$<2\xi_{\rm M}$ (top left panel) $|\Psi(x)|^{2}$ of the two lowest levels decay from both ends into the bulk of the superconducting nanowire. Such levels exhibit an spatial overlap, which is reduced as $L_{\rm S}$ increases (see bottom left panels). On the other hand, when the spatial overlap of the Majorana wavefunction is finite, the charge density $|\rho(x)|^{2}$ develop an uniform oscillation pattern, predicted to be associated to MBSs.\cite{Ben-Shach:PRB15}  As $L_{\rm S}$ increases,  $|\rho(x)|^{2}$ gets reduced and reaches zero  when $L_{\rm S}\gg2\xi_{\rm M}$ (bottom right panel in Fig.\,\ref{wavefunc}), signalling charge neutrality of the two lowest levels (MBSs).

\subsection{SNS junction}
\label{app1}
In order to complete the analysis given in the main text, in this part we provide additional calculations for SNS junctions. 

\begin{figure}
\centering
\includegraphics[width=.45\textwidth]{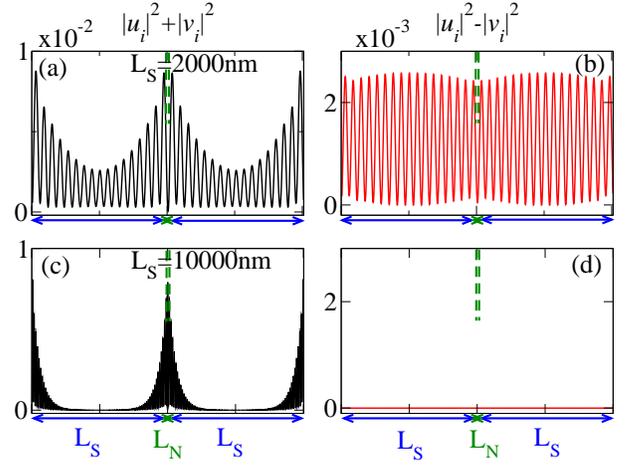} 
\caption{(Color online) (a,c) Wavefunction amplitude $|\Psi(x)|^{2}$ and (b,d) charge density $|\rho(x)|^{2}$ in short junctions for (a,b) $L_{\rm S}\leq 2\xi_{\rm M}$ and (a,b) $L_{\rm S}\gg 2\xi_{\rm M}$, corresponding to the two lowest levels (MBSs) in a topological superconducting nanowire. 
Parameters: $L_{\rm N}=20$\,nm, $\alpha_{R}=20$\,meVnm, $\mu_{\rm N}=\mu_{\rm S}=0.5$\,meV, $\Delta=0.25$\,meV and $B=2B_{\rm c}$.}
\label{wavefunc2}
\end{figure}

\begin{figure}
\centering
\includegraphics[width=.45\textwidth]{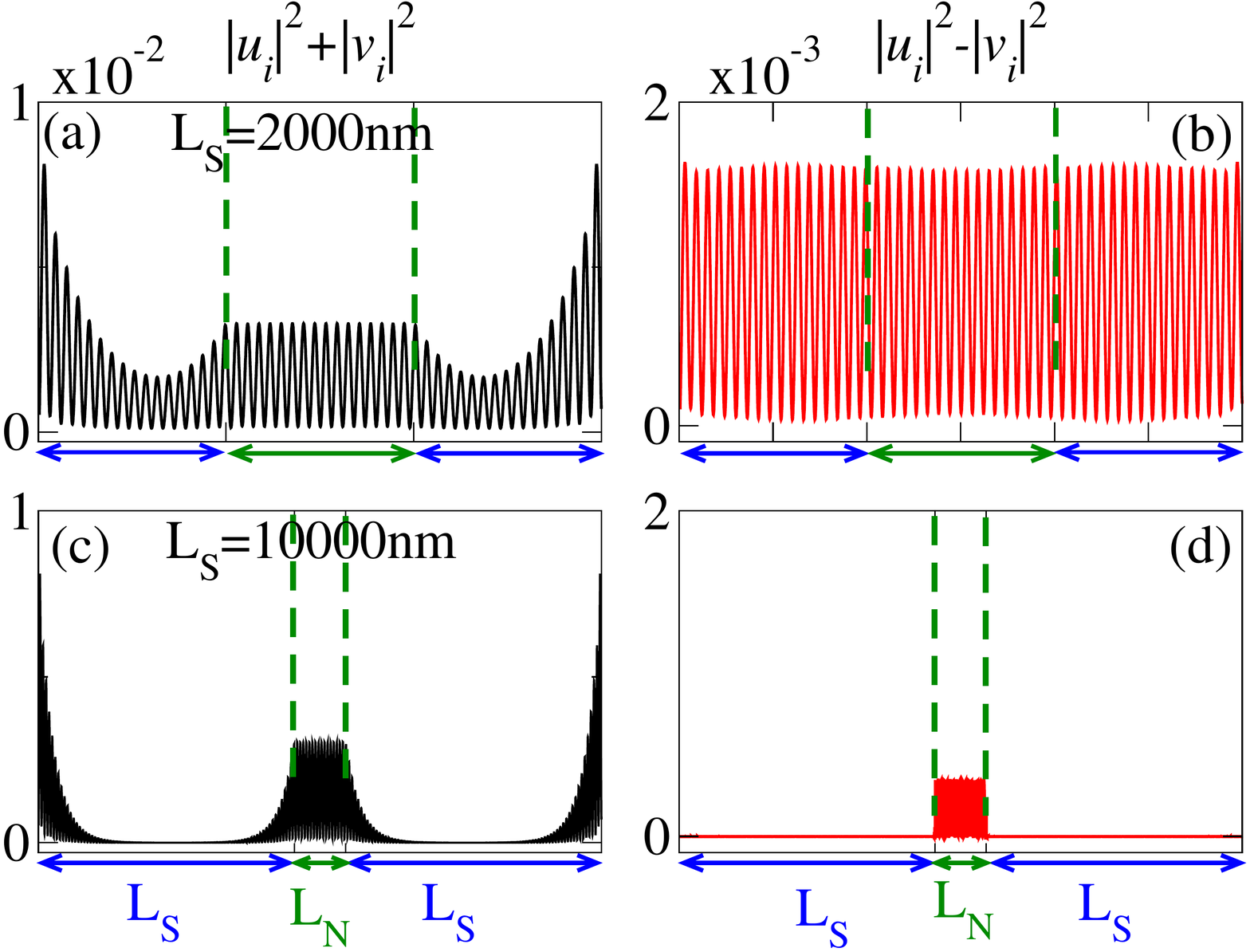} 
\caption{(Color online) (a,c) Wavefunction amplitude $|\Psi(x)|^{2}$ and (b,d) charge density $|\rho(x)|^{2}$ in long junctions  for (a,b) $L_{\rm S}\leq 2\xi_{\rm M}$ and (a,b) $L_{\rm S}\gg 2\xi_{\rm M}$, corresponding to the two lowest levels (MBSs) in a topological superconducting nanowire. 
Parameters: $L_{\rm N}=2000$\,nm, $\alpha_{R}=20$\,meVnm, $\mu_{\rm N}=\mu_{\rm S}=0.5$\,meV, $\Delta=0.25$\,meV and $B=2B_{\rm c}$.}
\label{wavefunc3}
\end{figure}

We firstly show in Figs.\,\ref{wavefunc2} and \ref{wavefunc3} the BdG wave functions amplitude $|\Psi(x)|^{2}$ and charge density $|\rho(x)|^{2}$ of the MBSs in short and long junctions. These calculations are obtained following similar analysis as in the previous section for the Rashba nanowire. Observe that MBSs are localized at the ends of the S regions, exhibiting a considerable overlap when $L_{\rm S}\leq2\xi_{\rm M}$ and a negligible one when $L_{\rm S}\gg2\xi_{\rm M}$. The associated charge density $|\rho(x)|$ exhibits uniform oscillations when the wave function overlap is finite, while it acquires zero value when the MBSs are located far apart, namely, for $L_{\rm S}\gg2\xi_{\rm M}$.

 \begin{figure}
\centering
\includegraphics[width=.45\textwidth]{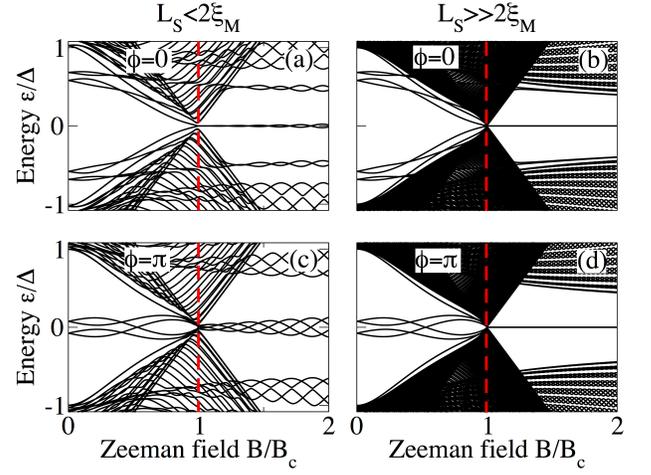} 
\caption{(Color online) Same as in Fig.\,\ref{SNSa} but for intermediate junction. 
Parameters: $L_{\rm N}=400$\,nm, $\alpha_{R}=20$\,meVnm, $\mu=0.5$\,meV and $\Delta=0.25$\,meV.}
\label{SNScx}
\end{figure}

Moreover, in  Fig.\,\ref{SNScx} we also present the  Zeeman dependent low-energy spectrum for intermediate SNS junctions.
Observe that as one increases the length of the normal region $L_{\rm N}$ additional levels, coming from the N region, appear within the induced gap. Also notice that 
the SOC minigap, being the separation between the four lowest levels and the rest of the energy spectrum, is reduced for the same reason. Notice that the low-energy spectrum in the trivial phase $B<B_{\rm c}$ is the same irrespective of $L_{\rm S}$, as seen in Fig.\,\ref{SNScx}(a-d). The topological phase, however, $B>B_{\rm c}$ is very sensitive to variations in $L_{\rm S}$. The same effect remains for short, intermediate and long junctions.
This has been demonstrated to have crucial influence on critical currents in short SNS junctions.\cite{Cayao17b}
\end{document}